\documentclass[twocolumn,tighten]{aastex63}

\usepackage{amsmath}
\usepackage{xspace}
\usepackage{multirow}
\usepackage{fancyapj}
\usepackage{mathtools}

\makeatletter
\def\restartappendixnumbering{\global\applettertrue
\setcounter{table}{0}
\setcounter{figure}{0}
\setcounter{equation}{0}
\def\thetable{\thesection\the\c@table}%
\renewcommand{\theHtable}{Supplement.\thetable}
\def\fnum@table{{\bf\tablename~\thetable}}%
\def\thefigure{\thesection\the\c@figure}%
\def\fnum@figure{{\bf\figurename~\thefigure}}%
}%
\makeatother

  {\list{}{\leftmargin=0.1in\rightmargin=0.1in}\item[]}%
  {\endlist}

\expandafter\def\csname editcolor1\endcsname{magenta}
\expandafter\def\csname editcolor2\endcsname{red}

\newcommand{\br}[1]{\ensuremath{\left[ #1 \right]} }
\newcommand{\rbr}[1]{\ensuremath{\left( #1 \right)} }

\newcommand{\E}[1]{\ensuremath{\times 10^{#1}} }

\newcommand{\msol}{\ensuremath{M_{\odot}}\xspace}
\newcommand{\tasc}{\ensuremath{T_\text{asc}}\xspace}

\newcommand{\cts}{\rm\,ct\per{s}\xspace}
\newcommand{\kev}{\rm\,keV\xspace}
\newcommand{\hz}{\rm\,Hz\xspace}
\newcommand{\ks}{\rm\,ks\xspace}
\newcommand{\km}{\rm\,km\xspace}

\newcommand{\kpc}{\rm\,kpc\xspace}
\newcommand{\s}{\rm\,s\xspace}
\newcommand{\ms}{\rm\,ms\xspace}
\newcommand{\hr}{\rm\,hr\xspace}
\newcommand{\per}[1]{\rm\,#1\ensuremath{^{-1}}\xspace}
\newcommand{\persq}[1]{\rm\,#1\ensuremath{^{-2}}\xspace}
\newcommand{\lumcgs}{\rm\,erg\per{s}\xspace}
\newcommand{\fluxcgs}{{\rm\,erg{\per{s}}{\persq{cm}}\xspace}}

\newcommand{\mucgs}{\ensuremath{{\rm\,G\,cm}^3}\xspace}

\newcommand{\nudot}{\dot\nu}

\newcommand{\rxte}{\textrm{RXTE}\xspace}
\newcommand{\xmm}{\textrm{XMM-Newton}\xspace}

\newcommand{\swift}{\textrm{Swift}\xspace}

\newcommand{\nicer}{\textrm{NICER}\xspace}
\newcommand{\integral}{\textrm{INTEGRAL}\xspace}

\newcommand{\src}{IGR~J17062\xspace}
\newcommand{\srcfull}{IGR~J17062--6143\xspace}

\begin{document}

\title{Long-term coherent timing of the accreting millisecond pulsar IGR J17062--6143}

\author{Peter Bult}
\affiliation{Department of Astronomy, University of Maryland,
  College Park, MD 20742, USA}
\affiliation{Astrophysics Science Division, 
  NASA's Goddard Space Flight Center, Greenbelt, MD 20771, USA}

\author{Tod E. Strohmayer} 
\affil{Astrophysics Science Division and Joint Space-Science Institute,
  NASA's Goddard Space Flight Center, Greenbelt, MD 20771, USA}

%\author[0000-0002-0380-0041]{C.~Malacaria}
\author{Christian Malacaria}
\affiliation{NASA Marshall Space Flight Center, NSSTC, 320 Sparkman Drive, 
  Huntsville, AL 35805, USA}\thanks{NASA Postdoctoral Fellow}
\affiliation{Universities Space Research Association, Science and Technology Institute, 
  320 Sparkman Drive, Huntsville, AL 35805, USA}

\author{Mason Ng}
\affil{MIT Kavli Institute for Astrophysics and Space Research, 
  Massachusetts Institute of Technology, Cambridge, MA 02139, USA}

%\author[0000-0002-9249-0515]{Zorawar Wadiasingh}
\author{Zorawar Wadiasingh}
\affiliation{Astrophysics Science Division, 
  NASA Goddard Space Flight Center, Greenbelt, MD 20771, USA}
\affiliation{Centre for Space Research, North-West University,
  Potchefstroom Campus, Private Bag X6001, Potchefstroom 2520, South Africa}
\affiliation{Universities Space Research Association (USRA), 
  Columbia, MD 21046, USA}
  
\begin{abstract}
  We report on a coherent timing analysis of the 163 Hz accreting
  millisecond X-ray pulsar IGR J17062--6143. Using data collected with the
  Neutron Star Interior Composition Explorer and XMM-Newton, we investigated
  the pulsar evolution over a timespan of four years. We obtained a unique
  phase-coherent timing solution for the stellar spin, finding the source to be
  spinning up at a rate of $(3.77\pm0.09)\E{-15}$ Hz/s. We further find that
  the $0.4-6$ keV pulse fraction varies gradually between 0.5\% and 2.5\%
  following a sinusoidal oscillation with a $1210\pm40$ day period. 
  Finally, we supplemented this analysis with an archival Rossi X-ray Timing Explorer
  observation, and obtained a phase coherent model for the binary orbit
  spanning 12 years, yielding an orbital period derivative measurement of
  $(8.4\pm2.0)\E{-12}$ s/s. This large orbital period derivative is inconsistent
  with a binary evolution that is dominated by gravitational wave emission, and 
  is suggestive of highly non-conservative mass transfer in the binary system.
\end{abstract}


\section{Introduction}
\label{sec:intro}

Accreting millisecond X-ray pulsars (AMXPs) are rapidly rotating neutron stars
in low-mass X-ray binary (LMXB) systems. They are characterized by the
fact that their millisecond rotation periods are directly apparent from their
X-ray emission in the form of coherent, highly sinusoidal periodic pulsations.
Such coherent pulsations are informative about the nature of the
neutron star and its accretion environment \citep[see, e.g.,][for
a review]{DiSalvo2020}. For instance: the precise shape of the pulse waveform
encodes information about the interior composition of the star
\citep{Poutanen2003}; tracking the pulse arrival times allows us to study the
evolution of the neutron star spin and binary orbit \citep{Hartman2008}, and
may be used to study accretion torque theory \citep{Psaltis1999b} and binary
evolution models \citep{Nelson2003}. 

A significant challenge to the study of AMXPs comes from the fact that they are
transient systems. Being powered by the accretion flow, the pulsations are only
visible during X-ray outburst. For most of the known AMXPs, these outbursts last
only a few days and are interspersed by several years or even decades of
quiescence, which makes it difficult to establish the long-term evolution of
the neutron star rotation. Additionally, the source luminosity usually changes
by orders of magnitude over the course of an outburst, implying large changes
in the mass accretion rate and thus in the accretion torque acting on the
neutron star.

Only three AMXPs have shown outbursts that last for several years:
HETE J1900.1--2455 \citep{Kaaret2006}, MAXI J0911--655 \citep{Sanna2017a}, and
IGR J17062--6143 \citep{Strohmayer2017}. The first two, however, are both
intermittent pulsars; after about a month into the outburst, the coherent
pulsations disappeared \citep{Patruno2012c, Sanna2017a, AtelBult19a}. This
leaves \srcfull (\src) as the only known AMXP to persistently exhibit accretion
powered pulsations over a timescale of decades.

The observational history of \src is somewhat unusual. The source was first
discovered with \integral in 2006 \citep{Churazov2007}, and estimated to have
entered outburst sometime in late 2005 or early 2006 \citep{AtelRemillard08}.
Since its discovery, the system has remained remarkably stable, showing a
persistent X-ray luminosity of $L_X \approx 6\E{35}\lumcgs$ with little change in its
intensity or spectral continuum \citep{Degenaar2017, Eijnden2018}.
The detection of a highly energetic intermediate duration Type I X-ray burst in 2012
identified the source as a neutron star \citep{Degenaar2012}. Similar
intermediate duration bursts were observed from \src in 2015 \citep{AtelNegoro15,
AtelIwakiri15} and 2020 \citep{AtelNishida20a} and have been attributed to helium
burning with an unusually deep ignition depth \citep{Keek2017}. Equating the measured
luminosity during the photospheric radius expansion phase of the 2015 X-ray
burst to the empirical Eddington luminosity, \citet{Keek2017} estimated the
distance to the source at $7.3\pm0.5\kpc$.
The 163\hz pulsations were not discovered until 2017 from the single archival 
\rxte observation collected for this source in 2008 \citep{Strohmayer2017}. 
The binary ephemeris was subsequently determined with \nicer
\citep{Strohmayer2018a}, establishing that the pulsar resides in a 38-minute
ultra-compact binary, as was suspected based on multi-wavelength modelling
of the accretion disk emission \citep{Hernandez2019}. Notably, \citet{Strohmayer2018a}
measured a binary mass function of $(9.12\pm0.02)\E{-8}\msol$, which is the
smallest among known stellar binaries and implies a minimum companion
star mass of $0.006\msol$ (assuming a 1.4\msol neutron star). 

Although \src is unique in the sense that it is the only AMXP
to persistently show pulsations over decades, the long-term evolution
of its pulse properties has not yet been investigated in detail. 
To that end, we have executed a multi-year monitoring campaign of \src using
\nicer, with the aim of measuring the orbital and spin evolution of this pulsar. In
this work we present a phase-coherent timing analysis of these observations. We
further supplement these observations with \xmm data collected in 2016 to
obtain a long-term timing solution spanning 4 years, and with the 2008 \rxte
observation to extend the orbital solution to 12 years.

\section{Observations and Data Processing}
\label{sec:data}
\subsection{NICER}
  Between 2017 August and 2020 August \nicer has observed \src for a total
  unfiltered exposure of $372$\,ks. Based on their spacing in time, these data
  are naturally grouped into ten distinct epochs. We list these groups and their
  respective ObsIDs in Table \ref{tab:data}.

  We processed all data using \textsc{nicerdas} version 7a, which is released
  with \textsc{heasoft} version 6.27.2. By default, this pipeline selects only
  those epochs with a pointing offset $<54\arcsec$, a bright Earth limb angle
  $>30\arcdeg$, a dark Earth limb angle $>15\arcdeg$, and which are outside of the South
  Atlantic Anomaly. It additionally screens for signs of increased background
  emission by retaining only those data epochs that were collected at times when the rate of
  reset triggers (undershoots) is $<200$ ct/s/detector and the rate of high
  energy events (overshoots) is $<1$ ct/s/detector and $<1.52 \times
  \textsc{cor\_sax} ^{-0.633}$, where \textsc{cor\_sax} gives the cut-off
  rigidity of the Earth's magnetic field. We find that these latter two
  criteria are often too conservative when applied to the observations of \src.
  Following \citet{Bult2020a}, we first smoothed the overshoot light curve
  using 5-second windows to reduce noise and then only retained those epochs
  when the absolute rate was less than $1.5$ ct/s/detector and increased the
  scaling factor of the \textsc{cor\_sax} expression filter from $1.52$ to
  $2.0$. The undershoot filter was only an issue in data groups 3,4,5, and 8.
  These data were all collected at times when the Sun angle was comparatively
  low ($<65\arcdeg$), which increases the optical load on the instrument and
  affects the low-energy background. We increased the undershoot rate filter
  for these data from 200 ct/s/detector to 400 ct/s/detector. 
  For all ObsIDs we compare the light curves obtained with default screening
  criteria to those obtained with our relaxed parameters to ensure no spurious
  features are introduced into the data selection. 

  Inspecting the light curves, we find that a number of the individual
  exposures in the third data group show near identical light curve profiles in
  which the overall count-rate drops linearly from about {40\cts} to ${0\cts}.$ These
  decays coincide with a decrease in the number of stars traced by the star
  tracker, and reflects an obstruction of the field of view rather than being of astrophysical
  origin\footnote{In each case these anomalies can be attributed to the ISS
  solar panels passing through the instrument's field of view.}. We reprocessed
  these particular ObsID with the additional requirement that the number of
  stars in the star tracker (\textsc{st\_stars}) be larger than $38$, which
  effectively removed all spurious trends. 

  After screening our data, we are left with $202\ks$ of clean exposure
  (as opposed to the $187\ks$ retained under standard screening
  criteria). We applied barycenter corrections to these cleaned data using the
  \swift/UVOT position of \citet{AtelRicci08} and the JPL-DE405 solar system
  ephemeris \citep{Standish1998}. All dates reported in this work are therefore
  in terms of Barycentric Dynamic Time (TDB).

\begin{table}[t]
    \centering
    \caption{%
        Data overview
        \label{tab:data}
    }
    \begin{tabular}{l h h l l l}
      \hline \hline
	  Group & Cycle & Epoch & ObsID Range & Date & Exposure \\
	  %~     & ~     & ~     & ~           & ~    & ( ks )   \\
      \tableline
	  1  & 1 & 1 & 10341001 [01 -- 07] & 2017-Aug & 17.8 \\
	  2  & 1 & 2 & 10341001 [08 -- 10] & 2017-Oct & \phantom{0}9.8 \\
	  3  & 1 & 3 & 10341001 [11 -- 18] & 2017-Nov & 16.5 \\
	  4  & 1 & 4 & 10341001 [19 -- 22] & 2017-Nov & \phantom{0}2.1 \\
	  5  & 1 & 5 & 10341001 [23 -- 27] & 2018-Jan & \phantom{0}2.1 \\
	  6  & 2 & 1 & 26010101 [01 -- 04] & 2019-Apr & 27.5 \\
	  7  & 2 & 2 & 26010102 [01 -- 04] & 2019-Jul & 23.6 \\
	  8  & 2 & 3 & 20341001 [01 -- 03] & 2019-Oct & 16.9 \\
	  9  & 3 & 1 & 30341001 [01 -- 12] & 2020-Jun & 25.9 \\
	  10 & 3 & 2 & 36120101 [01 -- 11] & 2020-Aug & 60.1 \\
      \tableline
    \end{tabular}
    \flushleft
    \tablecomments{The rightmost column lists the clean exposure in ks.}
\end{table}

\subsection{XMM-Newton}
\xmm observed \src on 2016 September 13-15 (MJD 57645; see \citet{Eijnden2018}
for a detailed analysis). This data was collected with the \textsc{epic-pn} camera
in timing mode, yielding a relative time resolution of $30\mu$s. The absolute
timing calibration of \xmm has an accuracy of $100\mu$s \citep{Rosen2020}, which
translates to a systematic phase uncertainty of 0.02 cycles for \src. Hence, these
data can be used to extend the coherent timing analysis of our \nicer observations.

We processed the \xmm data in SAS v18 using the latest version of the calibration
files. Standard screening filters were applied, i.e; we kept only those events with
photon energies in the $0.4-10\kev$ range, with $\textsc{pattern}\leq4$ and
screening $\textsc{flag} = 0$. We extracted the source event list from \textsc{rawx}
columns $[34:42]$, and obtained background events from \textsc{rawx} columns
$[6:14]$. The source light curve showed a mean count-rate of $18\cts$ over the $61\ks$
exposure, with a background contribution of $0.26\cts$. Finally, we used \textsc{barycen}
to apply barycentric corrections to our source data, based on the same source coordinates
and ephemerides used for \nicer.

\subsection{RXTE}
\rxte observed \src on 2008 May 3 (MJD 54589.0) for a total good time exposure
of $1.1\ks$ in the proportional counter array (PCA). The data was collected with
the PCA operating in event mode, using 64 energy channels and a time resolution of 1/8192\s. A
detailed analysis of these data is presented by \citet{Strohmayer2017}. Here,
we applied standard processing and screening methods to extract a photon
event list for this observation, and used \textsc{faxbary} to apply
barycentering corrections. The background rate was estimated using 
\textsc{pcabackest}.

\section{NICER Light curve}
\begin{figure*}[t]
  \centering
  \includegraphics[width=\linewidth]{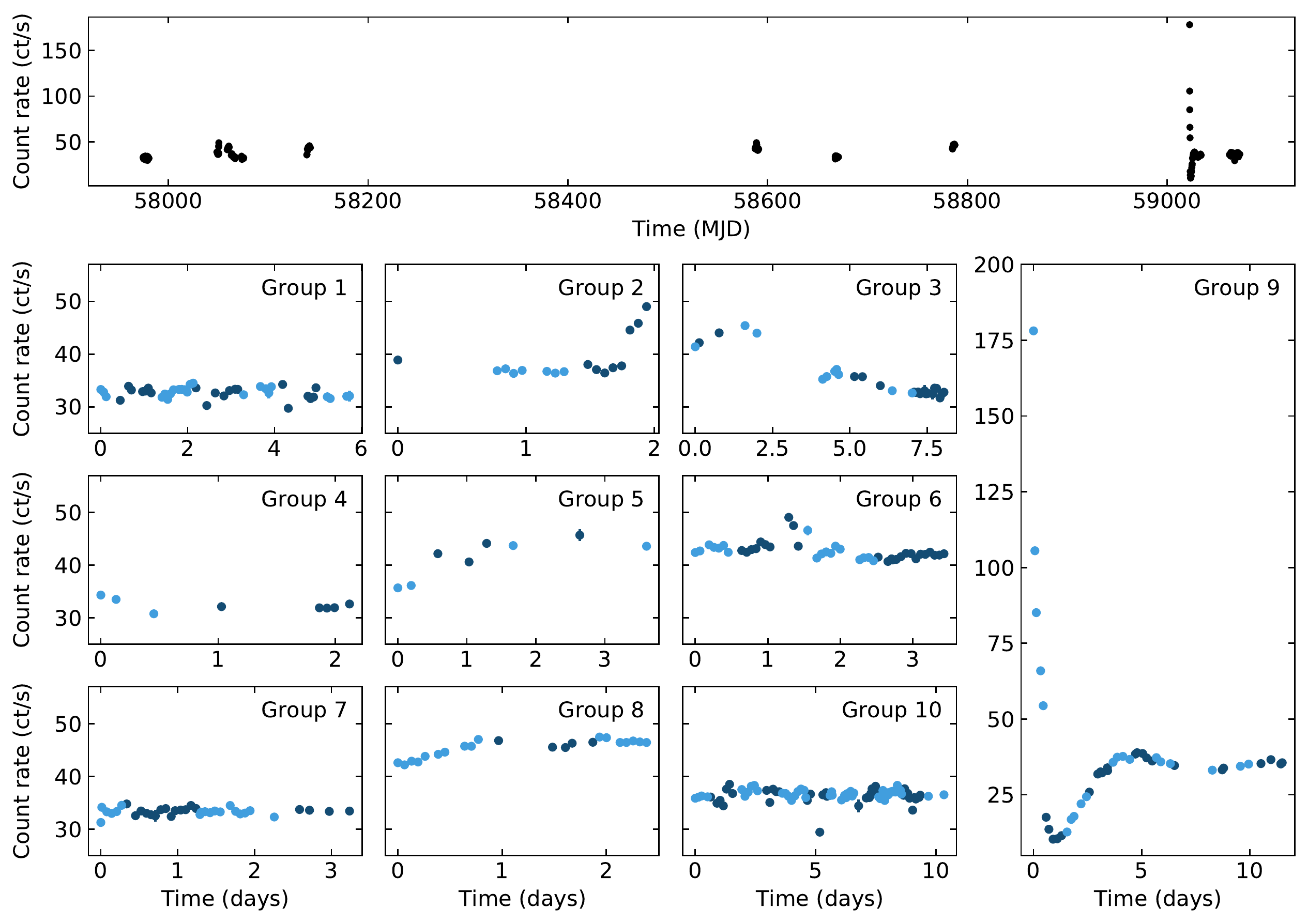}
  \caption{Background subtracted $0.3-6\kev$ \nicer light curve of \src. The
  top panel shows the complete light curve, with each point giving the average
  of a single continuous pointing. The bottom panels show the same data, but
  zoomed-in on the respective data groups as labeled. We plot the zoomed panels
  relative to the start time of each group, and alternate the point colors to
  indicate the even and odd ObsID numbers. The large swing in count-rate seen
  in group 9 is due to an intermediate duration X-ray burst from this source
  \citep{AtelNishida20a}.} \label{fig:lc}
\end{figure*}

  To investigate the source evolution over the span of our \nicer monitoring, we
  split the data into intervals representing continuous \nicer pointings. For
  each such pointing, we extracted a source spectrum and generated
  a background spectrum using version 6 of the 3c50 \nicer background model\footnote{%
    \url{https://heasarc.gsfc.nasa.gov/docs/nicer/tools/nicer_bkg_est_tools.html}
  } (Remillard et al., subm.). We find that the source emission dominates over the background between
  $0.4-6\kev$, and compute a background subtracted light curve in that energy
  range (Figure \ref{fig:lc}). We find that the source count-rate remains
  stable over the $3$ year baseline of our \nicer data, varying only slowly between
  approximately $30\cts$ and $50\cts$. Only in data group 9 do we see a large
  swing in source intensity. This data, however, was collected in response to
  an intermediate duration X-ray burst from this source \citep{AtelNishida20a}. The
  initial spike in count-rate is due to the cooling-tail of that X-ray burst,
  whereas the subsequent oscillation in source rate is similar to the system
  response after its 2015 X-ray burst \citep{Keek2017}.
  A detailed analysis of this event will be presented elsewhere.

\section{Coherent Timing}
\label{sec:method}

\subsection{Semi-coherent searches}
\label{sec:semi coherent}
Following the approach of \citet{Strohmayer2018a}, we separately search each data group
for the presence of 163\hz pulsations by optimizing the $Z_1^2$ score
\citep{Buccheri1983} as a function of spin frequency and the binary orbit's
time of ascending node, \tasc.  We adopt the ephemeris reported by
\citet{Strohmayer2018a} as our trial solution\footnote{Note that the spin
frequency reported in Table 1 of \citet{Strohmayer2018a} contains
a typographical error. We instead adopt the value quoted in the running text,
which we independently verified to be correct.}, and correct the photon arrival
times to remove the Doppler modulation imposed by the circular binary orbit. We
then compute the $Z_1^2$ score as
\begin{equation}
  Z_1^2 = \frac{2}{N} \br{ 
    \left( \sum_{j=1}^{N} \cos \varphi_j \right)^2
    + \left( \sum_{j=1}^{N} \sin \varphi_j \right)^2},
\end{equation}
with $\varphi_j = 2\pi\nu_i t_j$, where the $t_j$ give the list of $N$ photon
arrival times in a group, and $\nu_i$ is the test frequency. We evaluate this
score on a grid of trial frequencies spanning $\pm40\,\mu$Hz around the source
spin frequency. The window width was conservatively chosen such that we are
sensitive to a long-term spin frequency derivative smaller than
$4\E{-13}{\hz}\per{s}$ over the three year span between the first and last data
groups. The grid resolution is set to $1/(8T)$, where T is the duration of the
respective data group, that is, the time between the first and last photon
events in that group. For \tasc we adopt a grid spanning one orbital period
that we center on the middle of the respective data group's time interval. This
grid has a resolution of $25\s$, which is equivalent to $4\arcdeg$ orbital
longitude. 

We applied the $Z_1^2$ optimization method to all data groups, excluding group
9 which is dominated by the X-ray burst response, and groups 4 and 5, which do
not have sufficient exposure for this type of broadly defined search. 
Coherent 163\hz pulsations were recovered in all searched groups. We determined
uncertainties on the resulting spin frequency and \tasc measurements by
scanning the $Z_1^2$ space for the boundary interval where $\Delta Z_1^2 = 9$,
which corresponds to the $3\sigma$ confidence boundaries
\citep{Markwardt2002a}.
To further verify the validity of these parameter uncertainties, we simulated
$1000$ Poisson sampled event lists from the timing model while maintaining the
same count rates and good time intervals as our observations. Running the
$Z_1^2$ optimization method on these simulated data, we obtained parameter
uncertainties that were entirely consistent with those found through the $\Delta Z_1^2 = 9$
interval scan.
The resulting measurements and uncertainties are shown in Figure \ref{fig:semi
coherent}, with the detailed parameter values reported in the appendix. 

\begin{figure}[t]
  \centering
  \includegraphics[width=\linewidth]{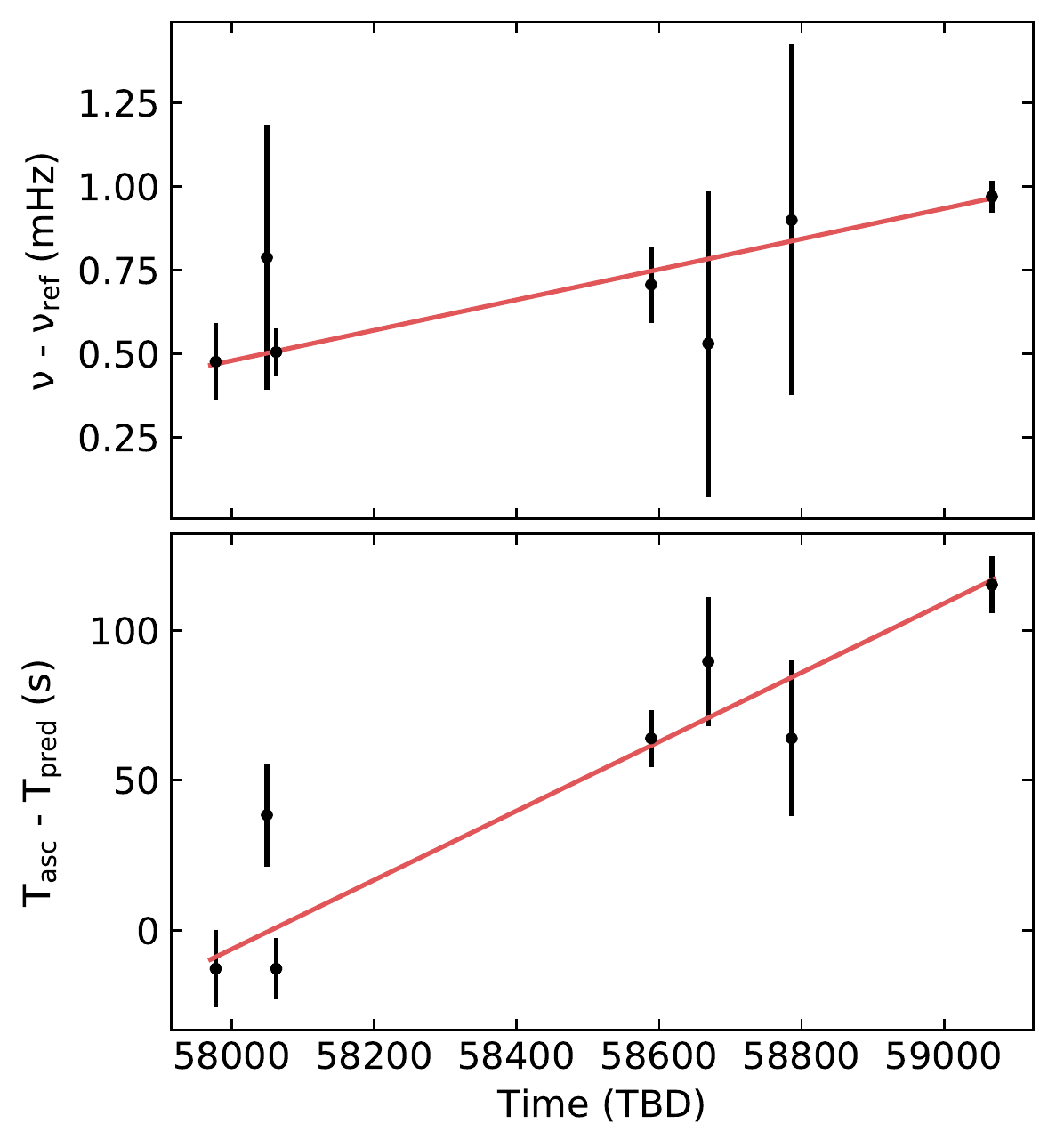}
  \caption{Individual spin frequency and \tasc measurements for each of the
  \nicer data groups listed in Table \ref{tab:data} (excluding groups 4, 5, and
  9). The top panel shows the frequency measurements in units of mHz relative
  to reference frequency $\nu_\text{ref} = 163.656110\hz$. The bottom panel
  shows the residual \tasc relative to the constant orbital period model
  of \citep{Strohmayer2018a}. For both panels the red line shows the best
  fit linear fit to the data.}
  \label{fig:semi coherent}
\end{figure}

Considering the pulse frequency measurements, we observe a modest linear 
drift across the data groups (Figure \ref{fig:semi coherent}, top panel).
To verify if this drift is statistically significant, we first fit the 
measurements using a constant frequency model. We obtain a best-fit $\chi^2$ of
$39$ for $6$ degrees of freedom, yielding a p-value of $7\E{-7}$. Hence,
this fit firmly rejects the constant spin-frequency hypothesis. Fitting the
measurements with a linear function instead, we obtain an acceptable fit, if
slightly under dispersed at a $\chi^2$ of $1.0$ for $5$ degrees of freedom.
The best-fit parameters of this linear fit give a spin-frequency measurement
of $\nu = 163.65611072 \pm 4\E{-8}\hz$ at a reference time of $58522.3$ TDB, with
a significantly detected long-term spin-up of $\dot\nu
= (5.3\pm0.9)\E{-15}{\hz}\per{s}$.

Considering the \tasc measurements, we observe a similar long-term evolution in
our measurements. In Figure \ref{fig:semi coherent} (bottom panel) we plot the
measured \tasc relative to the predicted values based on the constant orbital
period model of \citet{Strohmayer2018a}. These measurements are well described
using a linear function, yielding a best-fit $\chi^2$ of $8.4$ for $5$ degrees of freedom. 
The resulting best-fit orbital period is $P_b=2278.2110\s \pm 0.3\ms$, which is $3\ms$ 
larger than the orbital period reported in \citet{Strohmayer2018a} and well within
their $12\ms$ uncertainty region. Adding a quadratic term to the model, we can
further place a 95\% confidence upper limit on the long-term derivative of the
orbital period of $|\dot P_b| < 10^{-11}\s\per{s}$.

Using spin frequency and binary period parameters obtained from these linear models,
we refined the timing solution, and extrapolate the ephemeris to the epoch of the 
\xmm observation. We then selected the $0.4-6\kev$ events and corrected their
arrival times to the pulsar's reference frame using this predicted ephemeris.
Calculating the $Z_1^2$ score for this data, we obtain a score of $Z_1^2 = 18,$
which has an associated single trial p-value of $1\E{-4}$. Hence, the
pulsations are significantly detected in the \xmm observation at the
$4\sigma$ level. 

We also used the interim timing solution to consider \nicer data groups $4$,
$5$, and $9$. In each case we corrected the arrival times for the predicted
binary orbit and determined the $Z_1^2$ score. Significant pulsations
we recovered in groups $4$ (p-value of $3\E{-4}$; $3.6\sigma$) and
$5$ (p-value of $7\E{-6}$; $4.5\sigma$), but not in group 9 (p-value of 0.77).
Because the source luminosity varies drastically across data group $9$, we 
also searched each individual ObsID, however this did not yield
a pulse detection either. Finally, we attempted to suppress the influence of
the X-ray burst by processing the data in reverse order. That is, we first
searched the last ObsID for pulsations.  As none were detected, we added the
second-to-last and-so-forth. At no point in this process did the pulse
significance exceed $1\sigma$. In the absence of a detection, we calculate
a $95\%$ confidence level upper limit on the pulse fraction of $0.8\%$
if we include all ObsIDs in this group. We note, though, that this limit relaxes
to $1.1\%$ if exclude the first four ObsIDs, which show the largest swing
in source intensity (see Figure \ref{fig:lc}).

\subsection{Fully coherent pulse timing}
\label{sec:fully coherent}
A fully coherent description of the pulse signal requires that our timing
solution is precise enough to exactly predict the number of orbital revolutions and
stellar rotations that take place during the time intervals between the data
groups. Adopting the interim timing solution obtained in the previous section, we find
that the orbital phase is readily extrapolated across the time span of our
data (as is clear from Figure \ref{fig:semi coherent}). The stellar rotation,
on the other hand, requires more attention. The uncertainty in the predicted
pulse phase grows approximately as
\begin{equation}
  \sigma_\phi^2(t) = (\sigma_\nu t)^2 + (\frac{1}{2}\sigma_{\dot\nu} t^2)^2,
\end{equation}
where $\sigma_\nu$ gives the uncertainty on the spin frequency and
$\sigma_{\dot\nu}$ gives the uncertainty on its derivative. Using the
uncertainties obtained in Section \ref{sec:semi coherent}, we find that the
timing solution looses coherence ($\sigma_\phi$ exceeds $0.5$), when $t\approx200$ days.
Hence, we can coherently connect some data groups, but not all of them at once. To
further refine the timing solution, we therefore start by combining data groups
$2, 3,$ and $4$, which have the shortest time intervals between them. For each
group, we fold the data to a pulse profile, which we fit with a sinusoid to
measure the pulse phase residual relative to the model. The set of phase
residuals are then modeled using a quadratic function, so that we obtain refined measurements
for $\nu$ and $\nudot$, thus increasing the coherence time. By iteratively
adding the nearest data group to the analysis, we gradually converge to
a single timing solution that coherently describes all data. This approach
yielded a final spin frequency of $\nu=163.6561106613 \pm 1.0\E{-9}\hz$ and
a derivative of $\nudot = (3.73 \pm 0.07)\E{-15}{\hz}\per{s}$
measured relative to a spin epoch of $58522.3$ TDB.

\begin{figure}[t]
  \centering
  \includegraphics[width=\linewidth]{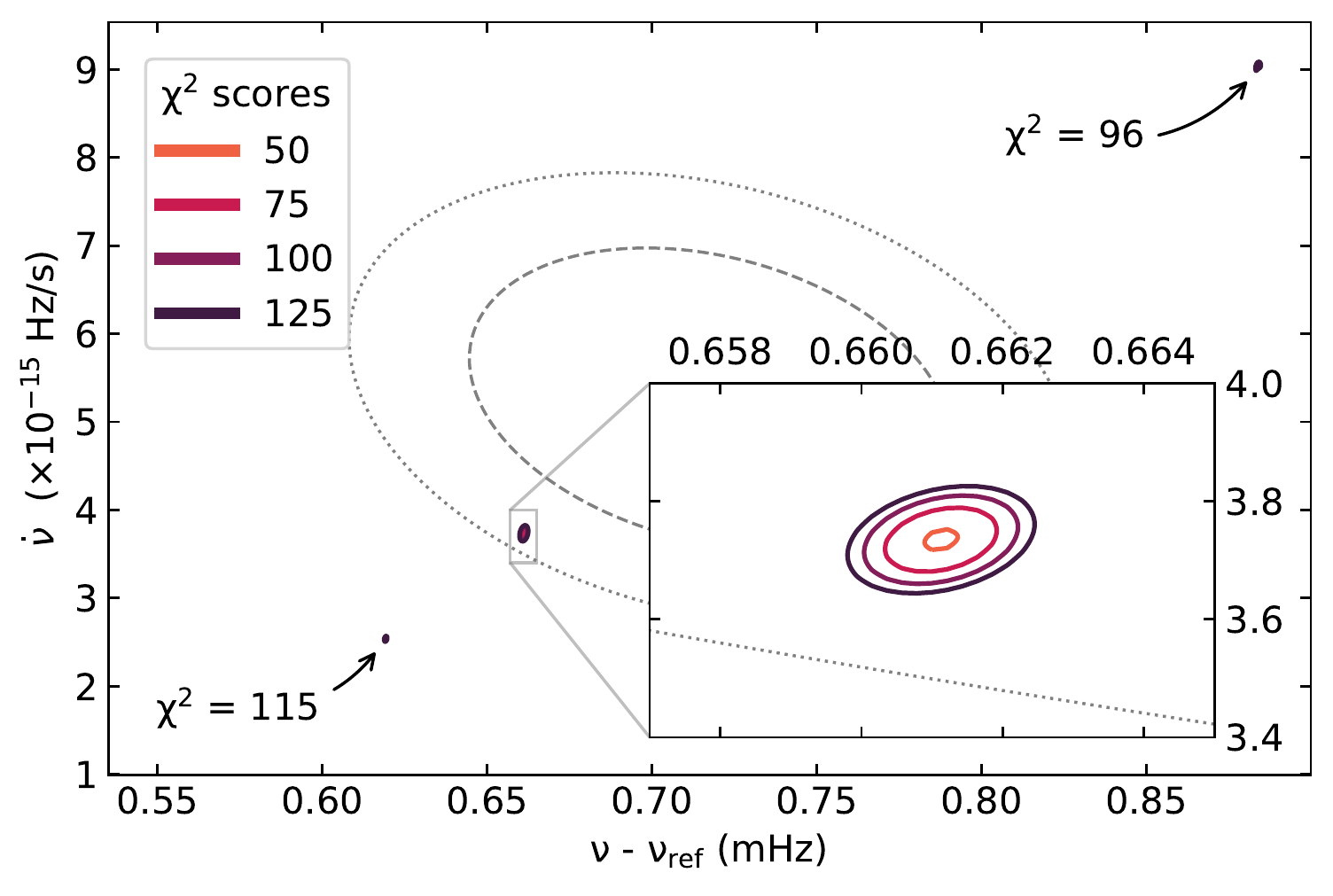}
  \caption{
    Phase-residual $\chi^2$-space as a function of spin frequency and spin
    frequency derivative. The grey dashed and dotted ellipses show
    the $2\sigma$ and  $3\sigma$ constraints obtained in Section 
    \ref{sec:semi coherent}. The four colored contours map out $\chi^2$ 
    levels of 50, 75, 100 and 125, which correspond to
    p-values of approximately $10^{-10}$, $10^{-15}$, $10^{-20}$, and $10^{-25}$,
    respectively.  Only one high-probability minimum is found (minimum
    $\chi^2=41$), with the inset showing a zoom-in of this region.
  }
  \label{fig:contours}
\end{figure}

While the iterative approach allowed us to determine a coherent solution for
the whole data set, this method is at risk of converging to a local minimum if
the timing solution is not unique. To verify the uniqueness of the solution, we
took a numerical approach. We scanned through the joint $5\sigma$ confidence
region in $\nu$ and $\nudot$ while taking steps of $10^{-10}\hz$ in frequency
space and steps of $10^{-18}{\hz}\per{s}$ in the derivative. For each trial
we compute the phase-residuals for the \xmm observation and each of the \nicer
data groups (excluding only group 9), and calculate the $\chi^2$ of this model.  As shown in Figure
\ref{fig:contours}, this $\chi^2$ scan yielded only one plausible timing
solution, which had the same $(\nu, \dot\nu)$ parameters found previously
through the iterative approach. 

We adopted the spin frequency and spin frequency derivative measurements associated
with the minimum in $\chi^2$ space and constructed the phase residuals. We then fit these
residuals jointly for the frequency evolution and the binary orbit parameters (orbital
period, projected semi-major axis, and \tasc). This fit yielded a slightly improved description
of the pulse arrival times and higher precision measurements of the orbital parameters.
We further attempted adding a second frequency derivative to the model, but this did not improve
the fit. Including orbital eccentricity or a binary period derivative was not required either.
Finally, optimizing the pulse phase residuals as a function of the source coordinate did not 
improve the residuals either, suggesting the source position is correct to
within a $90\%$ confidence uncertainty region of $0.1\arcsec$. The complete timing model
is listed in Table \ref{tab:timing solution}, while the resulting phase residuals
are shown in Figure \ref{fig:pulse panels}.

Using the refined timing solution, we can revisit the 2008 \rxte observation to
test for consistency. Extrapolating the spin frequency and \tasc to the \rxte
epoch ($54589.5$ TDB) and folding these data, we measure a pulse fraction
of about 7\%. Given that \citet{Strohmayer2017} report a pulse fraction of
$9\%$, this result indicates our timing solution is still suboptimal. We
therefore optimized the local timing solution by again determining the peak
$Z_1^2$ score as a function of the time of ascending node and the spin
frequency, finding $\tasc=54589.5510\pm0.0003$\,TDB and $\nu = 163.65612 \pm
5\E{-5}\hz$. This frequency measurement is consistent with the long-term
frequency solution, deviating only by $0.1\sigma$. The time of ascending node,
on the other hand, arrives 233 seconds later than predicted, which constitutes
a $3.3\sigma$ shift. Combining this measurement with the \tasc values obtained
from \nicer and \xmm, we find that the long-term drift in \tasc is incompatible
with a linear trend ($\chi^2/\text{dof}=25.9/7$), but well described by
a quadratic function ($\chi^2/\text{dof}=8.9/6$). The binary period derivative
obtained through this fit is $\dot P_b = (8.4\pm2.0)\E{-12}{\s}\per{s}$,
indicating that the binary orbit is expanding. 
The reference \tasc and orbital period obtained through this fit are consistent
with those found previously (to within $1\sigma$). 

\begin{table}
  \centering
  %\footnotesize
  %\hspace*{-3.4cm}
  \caption{\nicer timing solution \label{tab:timing solution}}
  \begin{tabular}{l l l}
    \hline \hline
    Parameter & Value & Uncertainty \\
    \tableline
    R.A. (J2000) & \phantom{--}17:06:16.29 & 0.1\arcsec \\
    Decl. (J2000) & --61:42:40.6 & 0.1\arcsec \\
    \tableline
    $\nu$ (Hz)                  & 163.6561106623 & 1.0\E{-9} \\
    $\dot\nu$ (Hz\per{s})       & 3.77\E{-15} & 9\E{-17} \\
    $|\ddot\nu|$ (Hz\persq{s})    & $<1.3\E{-24}$ \\
    Spin epoch              & 58522.3 \\
    $P_b$ (s)                 & 2278.21124   & 2\E{-5} \\
    $\dot P_b$ (s\per{s})     & 8.4\E{-12}   & 2\E{-12} \\
    Semi-major axis (lt-s)  & 0.003963     & 6\E{-6} \\
    \tasc (TDB)             & 58588.78464  & 5\E{-5} \\
    Eccentricity	    & $<0.03$ \\
    \tableline
  \end{tabular}
  \flushleft
  \tablecomments{Source coordinates were adopted from \citet{AtelRicci08}, with
  improved uncertainties obtained through the timing model. Upper limits
  are quoted at 95\% confidence.}
\end{table}

\subsection{Pulse variability}
\begin{figure}[t]
  \centering
  \includegraphics[width=\linewidth]{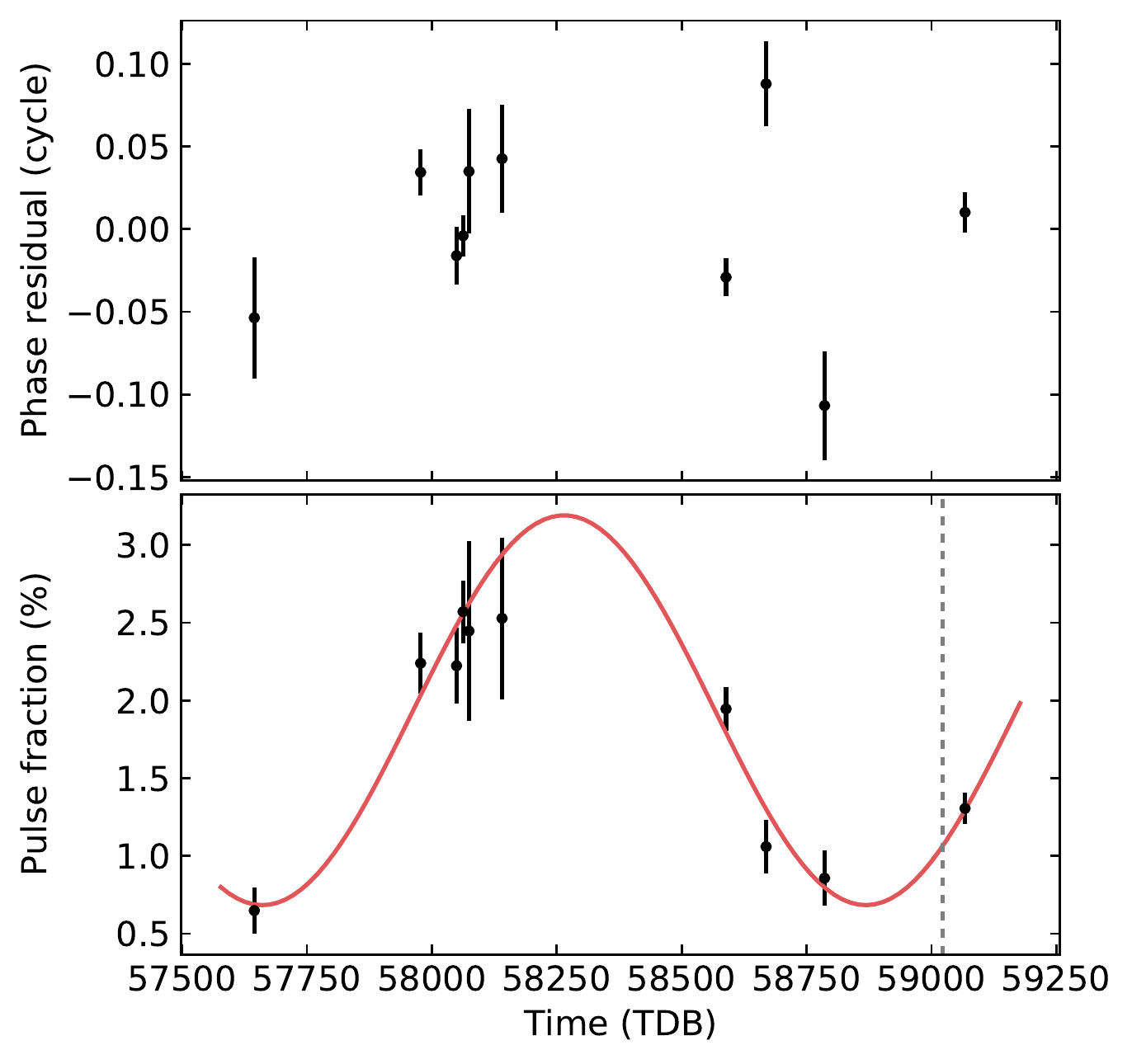}
  \caption{
    Pulse properties as a function of time. The top panel shows the pulse phase
    residuals relative to the timing model reported in Table \ref{tab:timing
    solution}. The bottom panel shows the pulse fraction of the fundamental
    pulsation, along with a best-fit sinusoidal model (red). The sinusoid has
    an amplitude of $(1.25\pm0.16)\%$ over a mean of $(1.94\pm0.08)\%$ and a period of
    $1210\pm40$ days. The vertical dashed line indicates the time of the 2020
    intermediate duration X-ray burst. 
  }
  \label{fig:pulse panels}
\end{figure}

Considering the pulse amplitudes across the different data groups, we observe 
clear evolution in pulse fraction over time. This trend could be well described
as a sinusoid ($\chi^2$/dof $= 6.4/6$), yielding an amplitude of $(1.25 \pm 0.16)\%$
over a mean pulse fraction of $(1.94\pm0.08)\%$, with a period measurement 
of $1210\pm40$ days, or about 3.3 years (see Figure \ref{fig:pulse panels}).
The $1-10\kev$ X-ray flux of these data groups does not exhibit any such
oscillation, instead showing a random 13\% rms scatter around a mean flux of
$\approx6\E{-11}\fluxcgs$.

\begin{figure}[t]
  \centering
  \includegraphics[width=\linewidth]{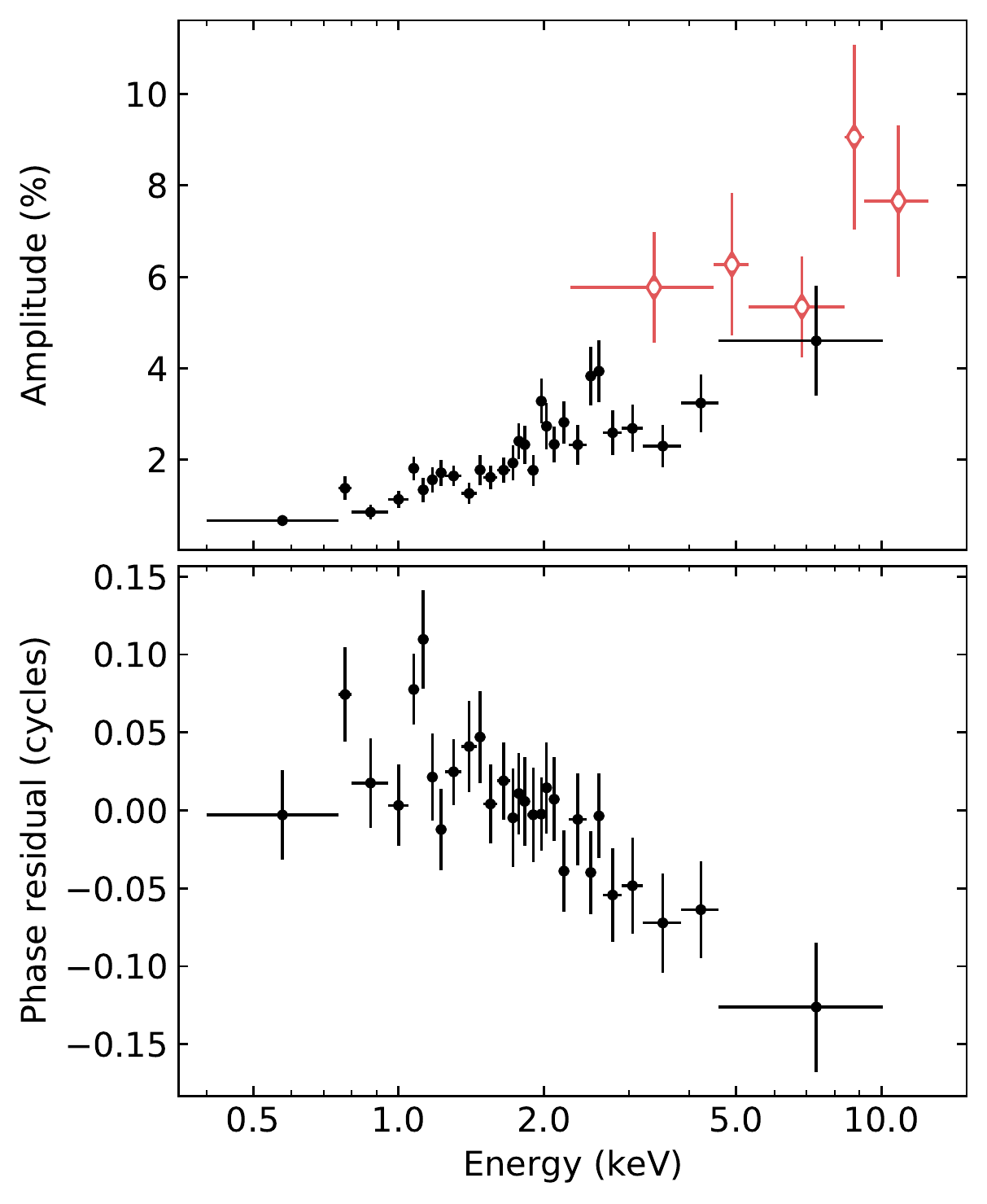}
  \caption{
    Pulse properties of \src as a function of photon energy, showing
    the pulse amplitude in the top panel and the pulse phase in the bottom
    panel. Solid black points show \nicer data and open red diamond show
    the \rxte measurements. The phases are expressed relative to the
    $0.4-6\kev$ energy band model reported in Table \ref{tab:timing solution}.
  }
  \label{fig:pulse energy}
\end{figure}

To investigate the energy dependence of the pulsations, we folded all \nicer data
to pulse profiles in several energy bins spanning $0.4-10\kev$. The bins
were constructed adaptively to be multiples of $0.5\kev$, such that each bin
contained at least $10^{4}$ counts and the pulsation was detected at $>5\sigma$
significance. As shown in Figure \ref{fig:pulse energy}, we find that the
time-averaged fractional pulse amplitude increases with energy to a peak of about 
$4.5\%$ between $5-10\kev$. For the pulse phase, we find that the pulsations have
a roughly constant phase below $2\kev$, but show an decreasing phase-lag toward
higher energies, i.e. the softer photons ($<2\kev$) tend to arrive after the 
harder photons ($>2\kev$).

For comparison, we also calculated the energy dependence of the pulse detection
in the 2008 \rxte observation. Using the local timing solution, we adaptively
binned individual event channels such that each bin contained a pulse detection
of at least $>5\sigma$. We find the pulsations could be resolved in five separate
energy bins, which are shown in Figure \ref{fig:pulse energy} using red
diamonds.

\section{Spectroscopy}

The pulse-phase averaged X-ray spectrum of \src has been studied in detail by
\citet{Degenaar2017}, \citet{Keek2017}, and \citet{Eijnden2018}. These studies
have consistently found that the continuum emission is well described by
a phenomenological model consisting of a single temperature blackbody with
a power law tail toward high energy, and a Gaussian emission line centered at
$1\kev$. 

In order to verify if this three component spectral model is appropriate for the
\nicer data as well, we extracted a $0.4-6.0\kev$ spectrum for each of the
\nicer data groups (excluding group 9). We model these
spectra in \textsc{xspec} v12.11 \citep{Arnaud1996} using the
T\"ubingen-Boulder interstellar absorption model \citep{Wilms2000}. We find
that the three component model gives a reasonable description for the continuum
emission in all groups, with best-fit spectral parameters that are consistent within
their respective uncertainties across the data groups. The model parameters (see the appendix
for detailed values) are broadly consistent with those reported by
\citet{Eijnden2018} based on 2015 \swift/XRT data and the 2016 \xmm
observation. The largest offset is found in the power law photon index, where
we measure a value of $\approx1.8$ versus the $2.0$ and $2.1$ reported by
\citet{Eijnden2018}. This difference is likely an effect of the energy
passband, however, as our \nicer spectra have a comparatively low upper bound
of $6\kev$, so that our sensitivity to the power law emission is quite poor. 

Having established that the phase averaged spectra do not vary substantially
across the different data groups, we proceeded with the phase-resolved
analysis.  Using the coherent timing solution obtained in Section
\ref{sec:fully coherent}, we assigned a rotational phase to each measured photon.
We then stacked all available data, and extracted $0.4-6.0\kev$ spectra in $8$
separate phase bins. We modelled these $8$ spectra jointly using the same three
component model used for the phase-averaged spectra. Initially, we let all
model parameters vary, keeping only the absorption column density tied across
the phase spectra. This fit yielded a good description of the continuum, with
all parameters in the same range as found from the phase-averaged analysis.
Considering the obtained spectral parameters as a function of phase bin, we
could observe an apparent oscillation in several spectral parameters (see Figure \ref{fig:phase 
resolved}). We modelled these variations using a sinusoidal model, 
requiring the sinusoidal amplitude to be greater than three times
its uncertainty for a significant detection. The modulation in the blackbody
temperature and power law flux were thus found to be significant, at
$4.3\sigma$ and $5.2\sigma$, respectively. Additionally, we found marginal
oscillations in the blackbody normalization ($2.1\sigma$), the power law photon
index ($2.7\sigma$), and the Gaussian line flux ($2.5\sigma$). The line energy
and width did not shown signs of phase dependence ($<1\sigma$).
The spectral and sinusoidal parameters are reported in Table \ref{tab:phase
resolved}.

\begin{figure}[t]
  \centering
  \includegraphics[width=\linewidth]{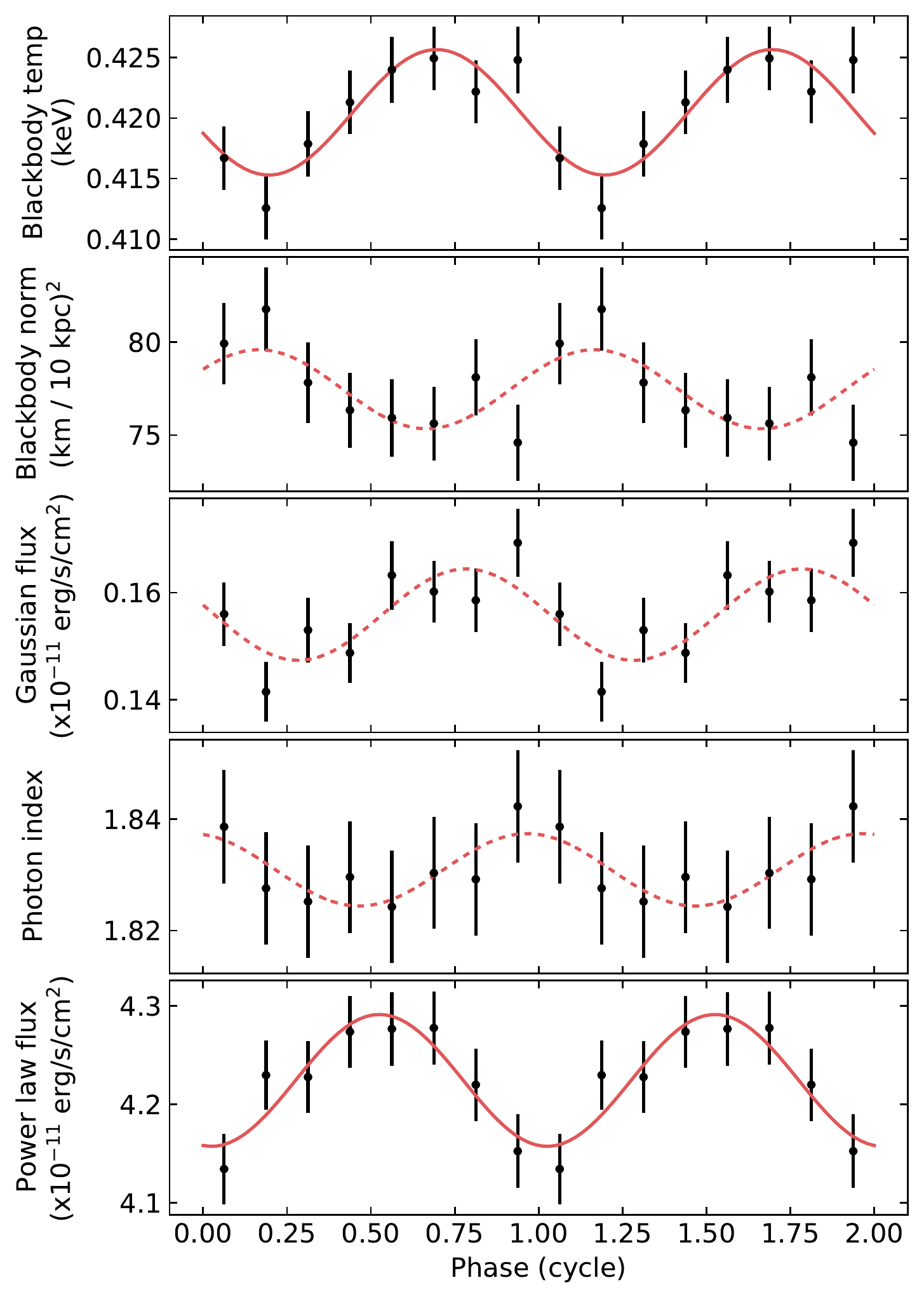}
  \caption{
    Phase resolved spectroscopic parameters of \src, showing from top to
    bottom: the blackbody temperature, the blackbody normalization, the
    Gaussian line flux, the power law photon index, and the power law flux. Both
    flux measurements are obtained for the $1-10\kev$ energy band. In each
    panel we show the respective spectral parameter across two rotation cycles
    (black), along with the best fit sinusoidal model (red). We use a solid
    line where the spectral parameter shows a significant pulsation ($>3\sigma$),
    and a dashed line where the parameter pulsation is marginal ($2-3\sigma$). The
    remaining spectral parameters were held tied across the phase bins, with
    all model parameters listed in Table \ref{tab:phase resolved}.
   }
  \label{fig:phase resolved}
\end{figure}

\begin{table}
  \centering
  \footnotesize
  \newcommand{\mc}[1]{\multicolumn{2}{l}{#1}}
  \caption{Phase-resolved spectroscopy \label{tab:phase resolved}}
  \begin{tabular}{l l l l}
    \hline\hline
    Component & Parameter & Value  \\
    \tableline
    Absorption & $N_H$           & \mc{$(0.110 \pm 0.003)\E{22}$ cm$^{-2}$} \\
    Gaussian   & line flux       & \mc{$(1.59 \pm 0.09)\E{-12}\fluxcgs$} \\
    Gaussian   & line energy     & \mc{$0.961 \pm 0.006 $ \kev}  \\
    Gaussian   & line sigma      & \mc{$0.182 \pm 0.009$ \kev } \\
    Blackbody & mean kT       & \mc{$0.421 \pm 0.002$ \kev } \\
    Blackbody & mean norm   & \mc{$77.5 \pm 1.2$ (km / 10 kpc)$^2$ } \\
    Power law & mean index & \mc{$1.831 \pm 0.003$} \\
    Power law & mean flux   & \mc{$(4.22 \pm 0.02)\E{-11}\fluxcgs$} \\
    \hline\hline
    Component & Parameter & Amplitude & Phase  \\
    \tableline
    Blackbody & kT    & $(1.3 \pm 0.5)\%$ & $0.38 \pm 0.06$ \\
    Blackbody & norm  & $(2.7 \pm 2.1)\%$  & $0.85 \pm 0.12$\\
    Gaussian  & flux  & $(5.5 \pm 3.6)\%$ & $0.47 \pm 0.11$ \\
    Power law & index & $(0.4 \pm 0.2)\%$ & $0.65 \pm 0.10$ \\
    Power law & flux  & $(1.6 \pm 0.5)\%$ & $0.21 \pm 0.05$ \\
    \tableline
  \end{tabular}
  \flushleft
  \tablecomments{Flux measurements correspond to the $1-10\kev$ band. Uncertainties
  are quoted at 90\% confidence.}
\end{table}

\section{Discussion}
\label{sec:discussion}

We have presented a timing analysis of the 163\hz pulsations of \src using
\nicer and \xmm observations spanning four years. We found that these data
could be described with a coherent timing solution, allowing us to measure the
positive spin frequency derivative. Additionally, we studied the time and energy
dependence of the pulsations, finding the pulse amplitude to vary slowly with
time, suggesting a years-long oscillation. Supplementing these data with a 2008
\rxte observation, we obtained a coherent model for the binary orbit spanning
12 years and a measurement of the binary period derivative. In the following 
we discuss the implications of these measurements. 

\subsection{Long-term spin up}
\label{sec:spin up}
We have measured the long-term spin frequency derivative of \src to be $\dot\nu
= (3.73\pm0.03)\E{-15}{\hz}\per{s}$, indicating the neutron star is slowly
spinning up under the influence of sustained mass accretion onto the star.
Positive pulse frequency derivatives have been previously reported for several
other AMXPs during outburst \citep{DiSalvo2020}, with magnitudes on the order
of $10^{-14}\sim10^{-13}{\hz}\per{s}$. In all cases these derivatives were
measured by modelling the quadratic drift in pulse phase. However,
flux-dependent timing noise can bias such phase measurements, making
it difficult to disentangle what part of the observed phase drift is intrinsic
to the pulsar rotation and what part is instead due to accretion driven
variability in the hot spot \citep{Patruno2009b, Bult2020a}. Our spin frequency
derivative measurement for \src does not suffer from this ambiguity: the spin
change is directly visible in frequency space.

In the conventional picture of accretion onto a magnetized star, the spin
evolution of the star is governed by the interaction between the stellar
magnetosphere and the inner accretion disk. The stellar magnetic field truncates
the accretion disk at $r_\text{m}$, the magnetospheric radius. Just outside $r_\text{m}$, the
field threads the disk leading to an exchange of angular momentum between the star and
the accretion flow that depends on the difference in their angular velocities. While
the precise torques acting on the star depend on the specifics of the interaction
model, the torque is generally expressed as 
\begin{equation} \label{eq:ntotal1}
  N_\text{total} = \dot M_\text{ns} \sqrt{G M_\text{ns} r_\text{m}} + N_\text{field},
\end{equation}
where $\dot M_\text{ns}$ is the mass accretion rate onto the neutron star, $G$
the gravitational constant, and $M_\text{ns}$ the mass of the neutron star. The
$N_\text{field}$ term expresses the torque exchange
between the magnetosphere and the inner accretion disk, and depends on specific
assumptions about the microphysics of the magnetic coupling. \citet{Rappaport2004}
studied the specific case of an accretion disk around a millisecond pulsar, and
derived
\begin{equation*}
  N_\text{field} = 
  \begin{dcases}
    \frac{\mu^2}{3 r_\text{c}^3} \br{\frac{2}{3} - 2\rbr{\frac{r_\text{c}}{r_\text{m}}}^{3/2} + \rbr{\frac{r_\text{c}}{r_\text{m}}}^3} & r_\text{m} < r_\text{c}, \\
    - \frac{\mu^2}{9 r_\text{c}^3} & r_\text{m} > r_\text{c}
  \end{dcases}
\end{equation*}
where $\mu$ is the magnetic dipole moment and
\begin{equation}
  r_\text{c} = \rbr{\frac{G M_\text{ns}}{\Omega_s^2}}^{1/3},
\end{equation}
the corotation radius, with $\Omega_s=2\pi\nu$ the angular velocity of the
neutron star. Adopting a $1.4\msol$ neutron star, we find that the corotation
radius of \src is $56\km$.

The measured spin frequency derivative gives us a measure of the total torque
acting on the star as $N_\text{total} = 2\pi I \dot \nu$,  where $I
= (0.5-2)\E{45}\,\text{g\,cm}^2$ gives the neutron star moment of inertia
\citep{Friedman1986, Steiner2015}. Substituting this expression into equation
\ref{eq:ntotal1}, we can place a lower limit on the magnetic dipole moment of
$\mu \geq 1\E{26}\mucgs$, with a magnetospheric radius of $r_\text{m} \geq
50\km$. We can obtain more precise measurements if we further assume that the
magnetospheric radius scales with the mass accretion rate as
\begin{equation} 
  r_\text{m}  = \xi \rbr{\frac{\mu^4}{G M_\text{ns} \dot M^2}}^{1/7}, 
\end{equation} 
with scaling factor $\xi \approx 0.5$ \citep{Long2005, Bessolaz2008, Zanni2013}.
Assuming the X-ray flux is a good tracer of the mass accretion rate, we adopt
the $0.3-79\kev$ luminosity measured by \citet{Eijnden2018} to find that
the accretion rate onto the neutron star is $\dot M_\text{ns} = 2.5\E{-11}
\msol\per{yr}$. Considering the energetics and recurrence time of the X-ray
bursts, \citet{Keek2017} derive a similar accretion rate, suggesting that this
estimate is quite robust.  Solving the set of equations gives $\mu
= 2.9\E{26}\mucgs$ with a truncation radius of $52\km$.  

Obviously the derived dipole moment is highly model dependent, as somewhat
different prescriptions of the disk-field interaction will produce slightly
different outcomes. There may also be additional torques acting on the neutron
star, such as the torque associated with a propeller mechanism
\citep{Illarionov1975}. There is reason to suspect that a propeller mechanism
might be active in \src; for one, the mass flow rate through the accretion disk
has been estimated to be $1.8\E{-10}\msol\per{yr}$ \citep{Hernandez2019}, about
on order of magnitude higher than the accretion rate onto the neutron star.
Additionally, high resolution X-ray spectroscopy \citep{Degenaar2017,
Eijnden2018} revealed narrow emission lines, which can be interpreted as the
blueshifted emission of an outflow. If we assume that \src is in a weak
propeller regime \citep{Romanova2005, Ustyugova2006}, then part of the accretion
flow is ejected by the propeller, and the remainder accretes onto the neutron
star. In this case, the material pressure exerted on the magnetosphere can be much
greater than inferred from the X-ray flux, while the ejection mechanism itself
applies a negative torque to the neutron star. The torque equation becomes
\begin{equation} \label{eq:ntotal2}
  N_\text{total} = \dot M \sqrt{G M_\text{ns} r_\text{m}} + N_\text{field} + N_\text{propeller},
\end{equation}
where 
\begin{equation}
  N_\text{propeller} = \dot M_\text{ej} \sqrt{G M_\text{ns} r_\text{m}},
\end{equation}
with $-\dot M_\text{ej} = \dot M_\text{disk} - \dot M_\text{ns} < 0$ expressing
the rate at which mass is being propelled out of the system. If we assume that
the magnetospheric radius depends on $\dot M_\text{disk}$, we can again
solve the equation to find $\mu = 6.3\E{26}\mucgs$ at a radius of
$46\km$.

Whether we include a propeller torque or not, we therefore find that the
observed spin-up of the neutron star is consistent with a stellar magnetic
field of $B \approx 5\E{8}\,G$ and an accretion disk that is truncated close to
the corotation radius. Here it is interesting to note that X-ray reflection
spectroscopy of the Fe K$\alpha$ emission line is subject to a degeneracy
between the inner truncation radius of the disk and the system inclination
\citep{Eijnden2018}. Hence, if we use the truncation radius obtained from the
torque analysis to break this degeneracy, the system inclination would
have to be $30\arcdeg\sim35\arcdeg$.

\subsection{Binary evolution}
By modelling the advancement in the time of ascending node passages between 2008
and 2020, we obtained a binary period derivative measurement of $\dot P_b
= (8.4\pm2.0)\E{-12}{\s}\per{s}$. Hence, we find that the binary orbit is expanding with
an evolutionary timescale of $P_b / \dot P_b = 8.6$\,Myr. 

The apparent binary evolution of \src is much faster than expected from theory.
Generally, the mass transfer in an ultra-compact binary is believed to be driven
by angular momentum losses through gravitational wave emission \citep{Kraft1962},
causing the binary orbit to gradually widen with time. The expected binary
period derivative follows as \citep{Rappaport1987, Verbunt1993, DiSalvo2008}
\begin{align}
  \dot P_b 
    &= 4.4\E{-14} \rbr{\frac{M_\text{ns} M_c}{\msol}} \rbr{\frac{M_\text{total}}{\msol}}^{-1/3}
    \nonumber \\ &\times
    \rbr{\frac{P_b}{1\hr}}^{-5/3}
    \frac{n - 1/3}{n+5/3 - 2q} ~\s\per{s},
\end{align}
with $M_c$ the mass of the companion star, $q=M_c/M_\text{ns}$ the binary mass
ratio, and $n$ the index of the companion star's mass-radius relation ($n=-1/3$
for a degenerate star). Adopting a $1.4\msol$ neutron star and a $30\arcdeg$ inclination (see Section
\ref{sec:spin up}),
we find a companion mass of $M_c = 0.011\msol$ from the mass function \citep{Strohmayer2018a},
and obtain an expected orbital period derivative for \src of
$10^{-15}\s\per{s}$, which is clearly inconsistent with our measurement.

Similarly rapid orbital evolution has been observed in a number  
of other AMXPs and LMXBs \citep{Patruno2017b, DiSalvo2020}. 
In a few particularly well sampled sources, most notably SAX J1808.4--3658
\citep{Bult2020a}, the advancing \tasc shows complex residuals, suggesting that
the orbital period evolution is modulated (quasi-) periodically over decades. In \src,
such modulation is not apparent in the data. As the period derivative
measurement is strongly influenced by the data gap between the 2008 \rxte and
2016 \xmm observations, however, we note that further monitoring is required to
assess the stability of the period change. 

Various models have been proposed to explain the anomalously rapid evolution
observed in some X-ray binaries, most prominently: donor mass loss, spin-orbit
coupling, and enhanced magnetic braking \citep[see, e.g.,][for in-depth
discussions]{DiSalvo2008, Burderi2009, Patruno2017b,Sanna2017c}. Each of these models, however, faces some
challenges. Spin-orbit coupling relies on the presence of a variable mass
quadrupole in the companion star \citep{Applegate1994, Richman1994}, however,
a low-mass companion likely does not have a sufficient energy budget for
this mechanism \citep{Brinkworth2006}, which is only exacerbated for \src,
which has the smallest known mass function among stellar binaries
\citep{Strohmayer2018a}. Enhanced magnetic braking might play a role
\citep{Justham2006, Ginzburg2020}, but only if the companion star can sustain
a $\sim1$\,kG magnetic field and is also losing mass through a stellar wind.
In the following, we therefore limit our discussion to just the mass loss
model. 

If we assume that mass transfer is non-conservative, then the matter being
ejected from the binary system can carry away additional orbital angular
momentum. Assuming that this mass ejection channel dominates the evolution
timescale, the binary period derivative is, to good approximation, given by the
relation \citep{Rappaport1987}
\begin{equation*}
  \frac{\dot P_b}{P_b} = -\frac{\dot M_c}{M_c},
\end{equation*}
so that the required mass loss rate in the companion star must be $\dot M_c
\approx 2\E{-9}\msol\per{yr}$. 

\begin{figure}[t]
  \centering
  \includegraphics[width=\linewidth]{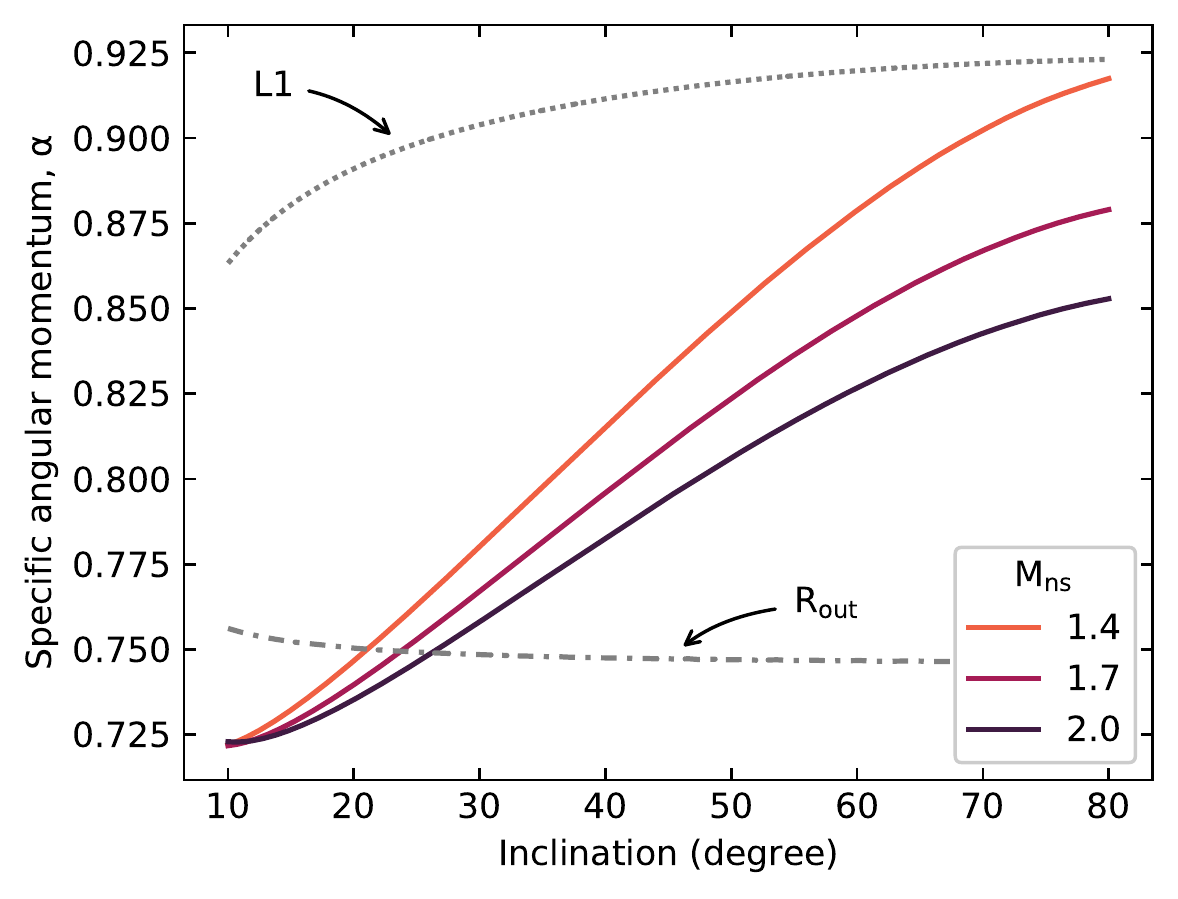}
  \caption{
    Specific angular momentum carried by the companion outflow, $\alpha$,
    expressed in units of the companion star's specific angular momentum as
    obtained by solving equation \ref{eq:mdot donor}. We show $\alpha$
    as a function of system inclination for three neutron star masses. We
    also show the specific orbital angular momentum of the inner Lagrange point
    (dotted line) and the outer edge of the accretion disk (dash-dotted),
    where the accretion disk radius was obtained from \citet{Hernandez2019}.
   }
  \label{fig:alpha}
\end{figure}

As we observe an ongoing outburst from \src, at least some of the matter lost
from the companion must be transferred to the neutron star. Following
\citet{Rappaport1982, Rappaport1983}, we can assume that a fraction $(1-\beta)$
of the matter lost by the companion is ejected from the system, carrying
specific angular momentum $\alpha$, expressed in units of the companion star's
orbital angular momentum.
The remainder of the mass loss is transferred to the neutron star/accretion disk,
so we can write $\dot M_\text{disk} = -\beta \dot M_c$, where the mass flow
rate through the disk has been observationally constrained to $\dot
M_\text{disk} = 1.8\E{-10}\msol\per{yr}$ \citep{Hernandez2019}.
Hence, about 90\% of the mass lost by the donor will be immediately ejected,
while the remaining 10\% will pass through the disk and move toward the neutron
star. Given that the mass accretion rate onto the neutron star is another order
of magnitude lower \citep{Keek2017, Eijnden2018, Strohmayer2018a}, it seems that
most of the matter flowing through the disk will eventually also be ejected.
This secondary outflow, however, likely occurs as a disk wind or a propeller,
meaning that the specific orbital angular momentum it carries is that of the
neutron star ($\alpha_\text{ns} \approx (a_1 / a_2)^2 = 0.01$) rather than that of
the companion star ($\alpha\approx1$). Hence, the impact of the second outflow
on the binary evolution is expected to be negligible. 
Under these assumptions, the mass loss rate of the companion is given as
\citep{Rappaport1983, Verbunt1993}
\begin{equation}
  \label{eq:mdot donor}
  \dot M_c = \frac{32 G^3}{5c^5}\rbr{\frac{4\pi^2}{G}}^{4/3} \rbr{\frac{M_\text{ns}}{P_b}}^{8/3}
  \frac{q^2}{(1+q)^{1/3} f_r},
\end{equation}
where
\begin{equation}
  f_r = \frac{5}{6} + \frac{n}{2} - \beta q - \frac{(1-\beta)(3\alpha+q)}{3(1+q)}.
\end{equation}
Equating this expression to the mass loss rate implied by the orbital period
derivative, we can solve for $\alpha$ as a function of just the neutron star mass
and system inclination (Figure \ref{fig:alpha}). We find $\alpha = 0.72-0.91$,
which corresponds to an ejection radius whose specific orbital angular
momentum lies somewhere between the specific orbital angular momentum carried
by the outer edge of the accretion disk ($\sim0.75$) and that of the inner
Lagrange point ($\sim0.9$).

What remains unaddressed, is what is powering this large mass loss rate. One
possibility is that the required energy is injected via the irradiation of the
donor by the luminosity generated by the neutron star. For \src, the accretion
luminosity is about $6\E{35}\lumcgs$. Assuming isotropic emission, the fraction
of radiation intercepted by the companion star is $f = (R_c/2a)^2 = 0.23\%$,
giving an irradiation rate of $\dot E_\text{irr} = 1.3\E{33}\lumcgs$. The
induced mass loss rate can be estimated as 
\begin{equation}
  \dot M = -\eta \dot E_\text{irr} \frac{R_c}{GM_c},
\end{equation}
so that $\eta\approx3\%$ of the impinging energy must be converted into 
the kinetic energy of the wind. In practice, this efficiency should be
considered a lower bound, as the companion star is likely shadowed by
the accretion disk, and any anisotropy in the accretion luminosity is
unlikely to beam the emission into the orbital plane. Whether or not this
efficiency is realistic is not clear, as the efficiency expected from theory
will depend greatly on the energy spectrum of the wind-driving radiation and 
the geometry of the companion star. Typically assumed efficiencies range
from 0.01-10\% \citep{Tavani1993, Justham2006, Ginzburg2020}, so it seems
that the efficiency required to drive a wind in \src is at least plausible,
if on the high side.

Finally, we note that \citet{Strohmayer2018a} derive a system inclination of
$19\arcdeg - 27.5\arcdeg$ based on the assumption that the binary evolves on
the timescale of gravitational wave emission. Clearly, this assumption cannot
be correct, so that the constraint on the inclination does not hold either.

\subsection{Phase resolved spectroscopy}
Consistent with previous spectral analyses of \src \citep{Degenaar2017,
Keek2017, Eijnden2018}, we found that the \nicer data were well described using
a phenomenological model consisting of an absorbed blackbody and power law,
with an additional Gaussian emission line centered at $1\kev$. Resolving the
energy spectra as a function of the pulsar rotational phase, we find a significant
pulse modulation in the blackbody temperature and the power law flux.
Additionally, we found marginal evidence ($2-3\sigma$) for modulation in the
blackbody normalization, the Gaussian line flux, and in the power law photon
index. 

One should be careful not to over interpret the results from phase-resolved 
spectroscopy. The adopted spectral model is entirely phenomenological,
and likely does not fully reflect the physical processes underlying the observed
emission. A more robust analysis of the phase-resolved spectroscopy requires
detailed waveform modelling \citep[see, e.g.,][]{Salmi2018}, which
is well beyond the scope of this work. With that caveat in mind, however, we note
that the significant temperature oscillation of the blackbody component does
support the interpretation that this thermal emission component is associated
with the stellar surface \citep{Eijnden2018, Hernandez2019}.

A potentially quite interesting aspect of the phase-resolved spectroscopy is
that there is some (weak) evidence that the Gaussian line flux is modulated at
the stellar rotation rate. The precise origin of this emission line is not
known, but it is likely with either Fe-L band emission, or possibly
with a Ne X line \citep{Degenaar2013}. High resolution spectroscopy indicates
the $1\kev$ feature can be resolved into several narrow lines
\citep{Degenaar2017, Eijnden2018}, and might be attributed to a collisionally
ionized plasma, possibly from a shock in the accretion column. Alternatively,
the line might be due to a blue-shifted emission line in an outflow, or from an
accretion disk reflection feature \citep{Degenaar2017, Eijnden2018}. A possible
phase dependence of the line flux would offer interesting context to this
debate. If the line originates in a shock, then one would indeed expect that
the line be modulated by the stellar rotation, in which case the precise phase
delay and waveform of the line flux should carry information on the precise
location of that shock. Alternatively, if the line is associated with an
outflow or a reflection feature, we can hypothesize that a phase dependence
arises indirectly from the beamed pulsar emission sweeping periodically over
the disk (or outflow), opening an opportunity for a tomographic reflection
study of the accretion flow \citep[see, e.g.,][]{Ingram2017}. 
Given the long-term stability of its pulsations, the data quality of the energy
resolved pulsations can in principle be improved with further observations.
Hence, we suggest that \src is an especially interesting target for continued
monitoring and detailed pulse waveform modelling.

\subsection{The variable pulse amplitude}
Considering the pulse amplitudes measured in each of the separate observing
epochs between 2016 and 2020, we found that the $0.4-6\kev$ pulse fraction of
\src ranges from $0.5\%$ to $2.5\%$ following a slow evolution with time. This
evolution could be well described by a simple sinusoidal oscillation, yielding
a period measurement of $1210\pm40$\,days. Because we only observed a little
over a single cycle, we could not determine if this apparent modulation is
maintained over decades, nor if the oscillation is periodic or quasi-periodic. 

The 2008 \rxte observation does not yield any useful constraints on the long-term 
pulse amplitude evolution. While the $2-12\kev$ pulse amplitude of the
\rxte observation has a much higher pulse fraction of $9.4\pm1.1\%$
\citep{Strohmayer2017}, we found that this measurement is strongly dependent on
photon energy (Figure \ref{fig:pulse energy}). At $2\kev$ the pulse fraction
drops to only $6\pm1\%$. If we compare this to the time-averaged pulse fraction
found with \nicer, we see that the measurements are consistent within the
statistical uncertainty. What little tension still remains between \nicer and
\rxte observations could be due to the time-dependence. If we extrapolate the
periodicity observed over the past four years, then the \rxte observation
coincides with a peak of the sinusoid and should be expected to yield a higher
pulse fraction than the time-averaged \nicer data. Alternatively, if the time
modulation of the pulse fraction is not strictly periodic, then the residual
tension could worsen, suggesting a long-term decay of the pulse fraction.
Continued monitoring of the pulse fraction oscillation may be able to resolve
this uncertainty by determining the stability of the period. In that context we
note that if the recent evolution holds, the next peak in pulse fraction should
occur around 2021 September, with the subsequent minimum occurring around 
2023 May.

The origin of the variable pulse amplitude is not clear at this time. Because
the pulsations of an AMXP are powered by the variable accretion flow, it is
generally expected that some of the accretion variability is visible in the
pulse waveforms as well. Indeed, the sample of known AMXPs shows ample evidence
for such a transfer mechanism. On rapid timescales (seconds), the stochastic
and quasi-periodic noise of the accretion disk has been shown to modulate the
pulse amplitude \citep{Menna2003, Bult2017b}. Pulse amplitudes also routinely
show variations on slower day-long timescales, which are generally understood
to be stochastically driven, and are obviously correlated with variations in
the source luminosity \citep{Chou2008}, spectroscopy \citep{Kajava2011}, or
temporal variability \citep{Bult2015}.
The slowly oscillating pulse fraction of \src appears to be of a very different
nature. Its timescale is much longer than any pulse variability seen in other
AMXPs, and notably, it does not correlate with changes in the pulse phase,
source luminosity, or any other observable of the accretion system. 
In particular, the lack of correlation between the pulse fraction and the X-ray flux
would suggest that the modulation is not driven by the mass accretion rate. 
Instead, it may originate in a geometric effect such as a periodic modulation of
the apparent size and/or orientation of the stellar hot spot. For instance, it
may be that the neutron star is slowly precessing \citep{Zimmerman1979, 
Alpar1985, Jones2001, Link2003}, although the precession period in AMXPs has 
been suggested to be on the order of minutes to days \citep{Chung2008}, and
presumably would impact the pulse phase as well. 

Intriguingly, the only other process in \src with a similar timescale is the
duty cycle of its intermediate duration X-ray bursts. Assuming no X-ray bursts were
missed, the three bursts observed in 2012, 2015, and 2020 give an average burst
recurrence time of $1460$ days, only somewhat larger than the apparent cycle
period observed in the pulse fraction. Both the 2015 and 2020 X-ray bursts
occurred near the minima of the pulse fraction. For this relation to extend to
the 2012 X-ray burst as well, however, the pulse amplitude oscillation would
have to be quasi-periodic, so continued monitoring should be able to establish
if these phenomena are truly connected. If, however, they are indeed related,
then we can speculate that the changes in the pulse fraction are tied to some
physics associated with the deep neutron star crust, rather than the accretion flow.

\subsection{A comparison to other long outbursts}
A remaining open question is why, after two decades in outburst, the pulsations
of \src are still visible. The two other AMXPs that exhibited a multi-year outburst
only showed pulsations for a limited time early in their outbursts, suggesting
that perhaps the decaying pulsations are a feature of prolonged accretion
episodes. We briefly discuss how these other two cases compare to \src.

The best studied case is that of HETE J1900.1--2455 (HETE J1900). This source
was first discovered in 2005 June \citep{vanderspek2005} and remained in outburst
until late 2015 \citep{Degenaar2017b}. Throughout most of its $\sim10$ yr
outburst, HETE J1900 maintained a high luminosity of about $\sim4\E{36}\lumcgs$
\citep{Falanga2007a, Papitto2013}, with significant intensity variability throughout
\citep{Patruno2017c, Degenaar2017}. Its $377\hz$ pulsations were persistently
visible for about 22 days \citep{Kaaret2006, Galloway2008b}, before becoming
intermittent and subsequently disappearing entirely \citep{Patruno2012c}.
This decay has been suggested to be due to magnetic field burial \citep{Cumming2008},
which is supported by the apparent exponential decay in the accretion torque derived
from the pulsar timing \citep{Patruno2012c}. A particularly strange aspect of 
HETE J1900 is that during the first two months of outburst its pulse amplitudes
appeared to be tied to the occurrences of X-ray bursts: following an X-ray
burst, the pulse fraction would abruptly increase and then steadily decay over
a $\sim10$ day timescale \citep{Galloway2007}.  

In \src the X-ray bursts might also be occurring when the pulse amplitude is at
a minimum, but we did not observe a similar abrupt jump in the pulse fraction
following the 2020 burst. The analysis of the \nicer epoch containing the X-ray
burst complicated by the impact of the X-ray burst cooling tail, though. While
the X-ray burst emission dominates, we would not expect to see accretion
powered pulsations \citep[see e.g.,][]{Watts2012}. The subsequent dip in source
intensity suggests that accretion onto the neutron star is inhibited, in which
case we might reasonably expect the pulsations to be likewise suppressed.
Hence, we would expect that the pulse signal is significantly variable during
this time. Unfortunately, however, we lack the sensitivity to test this theory.
We find an upper limit on the pulse amplitude of $0.8-1.1\%$, which is
comparable to the pulse fraction we would expect from interpolating the
long-term evolution. Hence, we cannot be certain if the pulsations suppressed
of if they are present at the expected rate. We can, however, rule out an
abrupt factor of $\gtrsim2$ jump in pulse fraction like the one observed in
HETE J1900 \citep{Galloway2007}. 

The second case is that of MAXI J0911--655 (MAXI J0911). First discovered in
2016 February \citep{Serino2016}, the outburst of this source is still ongoing.
Observations collected during the first month in outburst found $340\hz$
pulsations \citep{Sanna2017c}, but these pulsations were no longer visible in
observations collected at later times \citep{AtelBult19a}. In addition to the
longevity of its outburst, MAXI J0911 shares a couple of similarities with
\src: it has a 44 minutes ultra-compact binary orbit \citep{Sanna2017c}, and it
shows similarly rare but energetic intermediate duration X-ray bursts
\citep{AtelNakajima20a, AtelBult20a}. Where the sources differ, though, is in
their luminosity. Like HETE J1900, the luminosity of MAXI J0911 is estimated to
be $\sim5\E{36}\lumcgs$ \citep{Sanna2017c}, an order of magnitude larger than
the $\sim6\E{35}\lumcgs$ luminosity of \src. Hence, a simple explanation for
the persistently visible pulsations of \src might be rooted in the much lower
luminosity. That is, the magnetic screening process may simply be ineffective
at the much lower accretion rate of \src \citep{Cumming2001}, or operate on
much longer timescales.

\facilities{ADS, HEASARC, NICER}
\software{heasoft (v6.27.2), nicerdas (v7a)}

\acknowledgments
This work was supported by NASA through the \nicer mission and the Astrophysics
Explorers Program, and made use of data and software provided by the High
Energy Astrophysics Science Archive Research Center (HEASARC). PB further
acknowledges support from the NICER Guest Observer program (80NSSC21K0128) and
the CRESST II cooperative agreement (80GSFC17M0002).

\bibliographystyle{fancyapj}

\appendix
\restartappendixnumbering
\section{Semi-coherent search results}
Table \ref{tab:semi coherent} lists the detailed search results of the
per-epoch $Z_1^2$ pulse searches applied to the \nicer observations. 

\begin{table}[h]
  \caption{%
	Semi-coherent search results
	\label{tab:semi coherent}
  }
  \centering
  %\hspace*{-1.2cm}
  \begin{tabular}{c c c c}
    \hline \hline
    Group & $\Delta\nu$    & \tasc & $Z_1^2$ \\
    ~     & (mHz) & (TDB) & ~ \\
    \tableline
    1	&  $0.48 \pm 0.11$  &  $57977.43825 \pm 1.5\E{4}$  &  130.67 \\
    2	&  $0.8  \pm 0.4$   &  $58049.31841 \pm 2.0\E{4}$  &  \phantom{1}66.96 \\
    3	&  $0.51 \pm 0.07$  &  $58062.44916 \pm 1.2\E{4}$  &  153.39 \\
    6	&  $0.71 \pm 0.11$  &  $58588.78465 \pm 1.1\E{4}$  &  179.50 \\  
    7	&  $0.5  \pm 0.4$   &  $58668.99685 \pm 2.5\E{4}$  &  \phantom{1}40.28 \\
    8	&  $0.90 \pm 0.5$   &  $58786.01840 \pm 3.0\E{4}$  &  \phantom{1}22.40 \\
    10	&  $0.97 \pm 0.05$  &  $59067.23529 \pm 1.1\E{4}$  &  163.92 \\
    \tableline
  \end{tabular}
  \flushleft
  \tablecomments{%
    Spin frequency and time of ascending node measurements obtained through the $Z_1^2$ optimization
    search.  The frequencies are expressed relative to a reference frequency as $\Delta\nu = \nu - \nu_\text{ref}$,
    where $\nu_\text{ref} = 163.656110\hz$.
  }
\end{table}

\section{Phase-averaged spectroscopy}
Table \ref{tab:phase averaged} lists the detailed best-fit spectral parameters obtained
from the per-epoch spectroscopic analysis of out \nicer observations. 

\begin{table}[h]
  \caption{%
	Phase-averaged spectroscopic results
	\label{tab:phase averaged}
  }
  \centering
  \hspace*{-3.4cm}
  \begin{tabular}{c h c c c c c c c}
    \hline \hline
    Group & $N_H$ & $E_\text{line}$ & $\sigma_\text{line}$ & $F_\text{Gaussian}$ & BB Norm & BB Temp & $\Gamma$ & $F_\text{PL}$ \\
    ~     & (cm$^{-2}$) & (keV) & (keV) & ~ & (km / 10 kpc)$^2$ & (keV) & ~ & ~ \\
    \tableline
    1  & $0.107 \pm 0.002$ & $0.943 \pm 0.006$ & $0.210 \pm 0.007$ & $0.158 \pm 0.009$ & $79   \pm 4  $ & $0.396 \pm 0.004$ & $1.784 \pm 0.012$ & $3.75 \pm 0.04$ \\
    2  & $0.107 \pm -1.00$ & $0.952 \pm 0.006$ & $0.191 \pm 0.007$ & $0.175 \pm 0.010$ & $71   \pm 4  $ & $0.419 \pm 0.005$ & $1.747 \pm 0.013$ & $4.70 \pm 0.06$ \\
    3  & $0.107 \pm -1.00$ & $0.945 \pm 0.006$ & $0.198 \pm 0.008$ & $0.144 \pm 0.008$ & $68   \pm 3  $ & $0.421 \pm 0.004$ & $1.821 \pm 0.012$ & $3.86 \pm 0.05$ \\
    4  & $0.107 \pm -1.00$ & $0.919 \pm 0.021$ & $0.228 \pm 0.022$ & $0.178 \pm 0.025$ & $65   \pm 7  $ & $0.432 \pm 0.012$ & $1.804 \pm 0.030$ & $3.08 \pm 0.13$ \\
    5  & $0.107 \pm -1.00$ & $0.968 \pm 0.012$ & $0.167 \pm 0.015$ & $0.143 \pm 0.017$ & $82   \pm 6  $ & $0.424 \pm 0.008$ & $1.845 \pm 0.022$ & $4.10 \pm 0.12$ \\
    6  & $0.107 \pm -1.00$ & $0.979 \pm 0.005$ & $0.174 \pm 0.006$ & $0.148 \pm 0.007$ & $73   \pm 3  $ & $0.425 \pm 0.004$ & $1.885 \pm 0.010$ & $5.36 \pm 0.05$ \\
    7  & $0.107 \pm -1.00$ & $0.965 \pm 0.003$ & $0.170 \pm 0.004$ & $0.163 \pm 0.005$ & $80.9 \pm 1.9$ & $0.428 \pm 0.002$ & $1.820 \pm 0.010$ & $4.60 \pm 0.04$ \\
    8  & $0.107 \pm -1.00$ & $0.954 \pm 0.005$ & $0.195 \pm 0.006$ & $0.143 \pm 0.006$ & $74.7 \pm 2.1$ & $0.416 \pm 0.003$ & $1.867 \pm 0.011$ & $3.21 \pm 0.03$ \\
    10 & $0.107 \pm -1.00$ & $0.962 \pm 0.003$ & $0.181 \pm 0.003$ & $0.153 \pm 0.004$ & $80.9 \pm 1.5$ & $0.412 \pm 0.002$ & $1.826 \pm 0.010$ & $3.91 \pm 0.02$ \\
    \tableline
  \end{tabular}
  \flushleft
  \tablecomments{%
    The absorption column density is $N_H = (0.107 \pm 0.002)\E{22}$ cm$^{-2}$,
    and was tied across the different groups. The Gaussian flux ($F_\text{Gaussian}$) and
    power law flux ($F_\text{PL}$) are measured in the $1-10\kev$ band and
    expressed in units of $10^{-11}  \fluxcgs$.
  }
\end{table}

\end{document}